\documentclass[twocolumn, preprintnumbers ,amsmath, amssymb, superscriptaddress, longbibliography]{revtex4}

\usepackage{graphicx}
\usepackage{bm}% bold math
\usepackage{color}
\usepackage{amsmath}
\usepackage{amssymb}
\usepackage{graphicx}
\usepackage{subfigure}

\begin{document}

\title{Statistics of shared components in complex component systems}

\author{Andrea Mazzolini} 

\affiliation{Physics Department and INFN, University of Turin, via
  P. Giuria 1, 10125 Turin, Italy}

\author{Marco Gherardi}

\affiliation{Sorbonne Universit\'es, UPMC Univ Paris 06, UMR 7238,
  Computational and Quantitative Biology, 15 rue de l'\'{E}cole de
  M\'{e}decine Paris, France}
\affiliation{CNRS, UMR 7238, Paris, France}

\author{Michele Caselle}

\affiliation{Physics Department and INFN, University of Turin, via
  P. Giuria 1, 10125 Turin, Italy}

\author{Marco Cosentino Lagomarsino}

\affiliation{Sorbonne Universit\'es, UPMC Univ Paris 06, UMR 7238,
  Computational and Quantitative Biology, 15 rue de l'\'{E}cole de
  M\'{e}decine Paris, France}
\affiliation{CNRS, UMR 7238, Paris, France} 
\affiliation{FIRC Institute of Molecular Oncology (IFOM), 20139 Milan,
  Italy}

\author{Matteo Osella\footnote{To whom correspondence should be addressed.
 Email: mosella@to.infn.it}}

\affiliation{Physics Department and INFN, University of Turin, via
  P. Giuria 1, 10125 Turin, Italy}

\begin{abstract}
 
 Many complex systems are modular. Such systems can be represented as
 ``component systems'', i.e., sets of elementary components, such as
 LEGO bricks in LEGO sets. The bricks found in a LEGO set reflect a
 target architecture, which can be built following a set-specific list
 of instructions.  In other component systems, instead, the underlying
 functional design and constraints are not obvious a priori, and their
 detection is often a challenge of both scientific and practical
 importance, requiring a clear understanding of component statistics.
 Importantly, some quantitative invariants appear to  be common to many
 component systems, most notably a common broad distribution of
 component abundances, which often resembles the well-known Zipf's law.
 Such ``laws'' affect in a general and non-trivial way the component
 statistics, potentially hindering the identification of
 system-specific functional constraints or generative processes. Here,
 we specifically focus on the statistics of shared components, i.e.,
 the distribution of the number of components shared by different
 system-realizations, such as the common bricks found in different LEGO
 sets. To account for the effects of component heterogeneity, we
 consider a simple null model, which builds system-realizations by random 
 draws from a universe of possible components. Under
 general assumptions on abundance heterogeneity,  we provide analytical
 estimates of component occurrence, which quantify exhaustively the
 statistics of shared components. Surprisingly, this simple null model
 can positively explain important features of empirical
 component-occurrence distributions obtained from large-scale data on
 bacterial genomes, LEGO sets, and book chapters. Specific architectural
 features and functional constraints can be detected from occurrence
 patterns as deviations from these null predictions, as we show for the
 illustrative case of the ``core'' genome in bacteria.

\end{abstract}

\pacs{87.18.Wd, 89.75.Da}

\maketitle

\section{Introduction}

\label{sec:intro}

A large number of complex systems in very different contexts - ranging
from biology to linguistics, social sciences and technology - can be
broken down to clearly defined basic building blocks or components. 
For example, books are composed of words, genomes of genes, and many
technological systems are assemblies of simple modules. 
Once components are identified, a specific realization of a system (e.g.,
a specific book, a LEGO set, a genome) can be represented by its parts
list, which is the subset of the possible elementary components (e.g. words,
bricks, genes), with their abundances,  present in the realization.
We use the term ``component systems'' for empirical systems 
to which this general representation can be applied. 

Occurrence patterns of components across realizations are expected to
reveal relevant architectural constraints.  For example, the bricks
present in each LEGO set clearly reflect a target architecture that
can be built with them following the instruction booklet.  While for
LEGO sets the assembly instructions are provided by the seller, in
most component systems the architectural constraints are not obvious.
Inferring such constraints from the statistics of components may
answer important questions about the nature of a system.  For example,
it could reveal new clues about the complex combination of selective
pressure and random events that shaped the functional composition of
extant genomes. 
Even in those cases where the architecture is partially or even fully
known and the instruction manual is available, the statistics of
components may help us distill some general principles
characterizing a given class of component systems, in some cases
revealing basic features of the underlying generative processes.

In order to perform detection of system-dependent features 
from patterns of shared components,  
we need to have a clear idea of the general behavior of 
component systems even in absence of functional constraints on 
the presence/absence of specific classes of components.   
This is by itself a challenging task, as such
systems show a large degree of non-trivial universal 
properties~\cite{Altmann2016,Koonin2011,lagomarsino2009universal} 
that could in principle affect the occurrence statistics. 
Indeed, several notable quantitative laws can be identified in the
composition of component systems of very different nature.  
This is well known, e.g., in linguistics, where the notorious ``Zipf's
law''~\cite{Zipf1935} describing the word frequency distribution (or
its equivalent rank plot) in a linguistic corpus has been the subject
of extensive investigations~\cite{zanette2005dynamics,Newman2005,
lu2010zipf,Eliazar2011,Font2015}.  In this context, the existence
of quantitative ``universal'' laws may in principle provide insights
on the cognitive mechanisms of text production, and can have practical
applications in data mining and data search
techniques~\cite{Altmann2016}.
Analogously, for genomes across the whole tree of life, the number of
genes in different evolutionary families is power-law distributed, a
discovery that represents one of the first examples of ``laws'' of the
genome sequencing era~\cite{Huynen1998,Koonin2011}. 
Such heterogeneous usage of the different basic components, often 
resulting in an approximately power-law distribution of their
frequencies, can be seen as a hallmark of the complexity 
of component systems~\cite{Newman2005}.

A large body of theoretical work addresses the origins of this
heterogeneity.  Several models have emerged in different areas of
science, with context-specific ingredients. For example, stochastic
processes based on gene duplication, deletion and innovation have been
proposed as simple evolutionary models of genome evolution at the
basis of the observed heterogeneous component usage~\cite{Qian2001,
karev2002birth, karev2003simple, lagomarsino2009universal}. On the
other hand, specific communication optimization principles
~\cite{mandelbrot1953informational, i2003least} and stochastic models
for text generation~\cite{simon1955class, zanette2005dynamics,
Gerlach2013} have been invoked to explain the emergence of Zipf's
law in natural language. In many, but not all, of these models a
preferential attachment principle is at the origin of the emergence of
the power-law distribution of component frequencies.
More importantly, the ubiquity of this emergent behavior raises
the question of whether (and to what extent) empirical laws like
Zipf's law are pervasive statistical patterns that transcend
system-specific mechanisms~\cite{Koonin2011,Baek2011}.  In this
spirit, the analysis of radically different systems can help the
discovery of patterns that descend from pure statistical effects or
general principles~\cite{Baek2011,pang2013universal}.

Here, we analyze empirical data from three very different component
systems from linguistics (book chapters), genomics (protein domain
families in sequenced genomes) and technology (LEGO toys) and we look
for general statistical \emph{consequences} of their heterogeneous
frequency distributions.
The different data sources considered here reasonably do not share any
generative mechanisms, nor are they expected to share the same type of
constraints, selection criteria or optimization principles. However,
the frequency of their components is heterogeneous and they all obey
laws that are similar to Zipf's.

The marginal statistics that we concentrate on is the fraction of
components that are shared among a certain number of realizations. For
example, the fraction of LEGO bricks with the same shape found in a
given fraction of unequal LEGO boxes.
In genomics, this is the so-called ``gene-frequency distribution'',
which was shown to follow a U-shape at several taxonomic
levels~\cite{touchon2009organised,Koonin2008,haegeman2012neutral}.
A U-shape of this distribution of shared components indicates that
there is a set of ``core'' components that are common to most
realizations, as well as an enriched set of realization-specific
components. This histogram also decays approximately as a power law
for rare components, both in genomic data and in technological
systems~\cite{pang2013universal}.
In evolutionary genomics, the origins of this pattern are the focus of
a lively debate.  The pattern has been rationalized theoretically by
neutral or selective population dynamics
models~\cite{haegeman2012neutral,lobkovsky2013gene,baumdicker2012infinitely,
collins2012testing}, or as a consequence of functional dependencies
among different components~\cite{pang2013universal}.
For component systems outside of genomics, the distribution of shared
components remains under-explored, and is typically neglected by the
current debate, for example in linguistics~\cite{Altmann2016}.

Using theoretical calculations based on random sampling of components
(with replacement) from their overall frequencies (estimated by their
total abundance across empirical realizations), we show that a
distribution of shared components with a power-law behavior is a
general feature of component systems not only with Zipf-like component
frequency distributions, but also for general power-laws and
exponential decay of the overall component frequencies.  In other
words, a U-shaped distribution of shared components can naturally
emerge in component systems with a heterogeneous component usage
(which is often the case empirically).
Importantly, we quantitatively identify the general features of the
system leading to a U-shaped distribution of shared components, a
given core size, and a specific decay of the realization-specific bulk
of this distribution.

\section{Data}
\label{data}
\subsection{Data sources}

\paragraph{Genomes.} We used the superfamily classification of
protein domains from the SUPERFAMILY
database~\cite{wilson2007superfamily} considering a set of $R = 1061$
prokaryotic genomes (``realizations'') and a total number of different
families $N = 1531$ (``components'').
Protein domain families are the basic modular topologies of folded
proteins~\cite{Orengo2005}.  Different domains of the same family can
be found in each genome in the same or different proteins.
As a functional annotation of protein domains in SUPERFAMILY, we
considered the SCOP annotations mapped into 7 general function
categories, as developed by C.~Vogel~\cite{Vogel2006}.

\paragraph{LEGO sets.} The composition in bricks of several LEGO sets
($R = 2820$) can be freely downloaded from ``http://rebrickable.com''.
We excluded from the analysis LEGO sets belonging to the category of
``LEGO Technic'' since, by construction, they share a very small
number of bricks with the classic LEGO toys. Similarly, we did not
consider LEGO sets with less than 80 components or belonging to the
categories ``Educational and Dacta'' and ``Supplemental'' in order to
exclude sets that are actually collections of spare parts or
additional bricks for other sets.

\paragraph{Texts.}
The analyzed linguistic corpus is composed by $R = 1721$ book chapters
(realizations) of several English books randomly chosen from the most
popular ones in the database ``http://www.gutenberg.org''.  We defined
chapters as realizations, instead of entire books, to obtain a corpus
with a range of sizes (total number of components per realization)
comparable to the one of genomes and LEGO toys (Figure S1).  The
complete list of books considered is reported in Table~S1 of the
Supplemental Material (SM).  The elementary components are defined as
the words regardless of capitalization (e.g.  ``We'' and ``we'' are
considered as the same component).

\subsection{Data structure:  Matrix representation of component
  systems}

\begin{figure}[h!]
\includegraphics[width=0.47\textwidth]{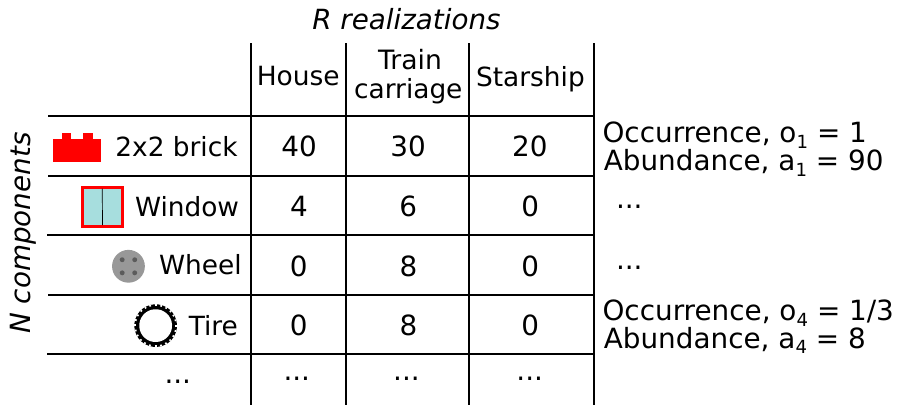}
\caption{\textbf{Matrix representation of complex component systems.}
  (a) Each column is a realization (e.g., a LEGO set, a genome or a
  book chapter) and each row is a component type (e.g., a LEGO
  brick, a protein domain family, a word).  The element $n_{ij}$ represents
  the abundance of component $i$ in realization $j$. The
  frequency $f_i$ of component $i$ is given by its total abundance
  ($90$ for the red brick, $8$ for the tire) divided by the total
  number of components in the system.  The occurrence $o_i$ of
  component $i$ is the fraction of realizations (toys in the example)
   in which there is at least one token of $i$ ($1$ for
  the red brick, $\frac{1}{3}$ for the tire).  }
\label{fig1}
\end{figure} 

A set of empirical realizations of a component system can be naturally
described as a matrix $\lbrace n_{ij} \rbrace$ defined such that the
entry $n_{ij}$ represents the abundance of the component $i$ ($i = 1,
\ldots N$) in the realization $j$ ($j = 1, \ldots, R$). Thus, each
realization (a literary text, a LEGO set or a prokaryotic genome), is
represented as a matrix column (Figure~\ref{fig1}).  Some key
observables can be easily defined using this representation.  First,
the total abundance $a_i$ of the component $i$ in the whole ensemble
is defined by summing over all
realizations $a_i = \sum_j n_{ij}$. The normalized abundance represents 
the component frequency  $f_i = \frac{a_i}{\sum_i  a_i}$.  \\
The ``component occurrence'' $o_i$ is instead defined as the fraction
of realizations in which the component is found, thus $o_i =
\frac{1}{R} \sum_j (1 - \delta_{n_{ij}, 0})$.  \\
Two other crucial quantities are: the total number $N$ of different
components in the system, which is essentially the number of 
bricks of different shape or the vocabulary, and the size of a realization $j$, defined
as the total number of its components $M_j = \sum_in_{ij}$.

\section{Results}

\subsection{Component frequency distribution and distribution of
  shared components show general features across systems}

This section illustrates two empirical laws in the analyzed datasets
(LEGO toys, bacterial genomes, and literary texts).
We first consider the component frequencies in the whole universe of
available realizations of a given system, which is essentially the
generalized Zipf's law~\cite{Newman2005} for the three systems.
Figure~\ref{fig2} shows the rank plots of these component frequencies.
The three data sets share a power-law behavior for components with
high frequencies (low rank), with an exponent close to 1 as in the
classic Zipf's law~\cite{Zipf1935}, and a faster decay at higher ranks
(components with low frequency).  This double-scaling behavior has
been recently observed in the context of
linguistics~\cite{Gerlach2013}. In evolutionary genomics, the gene
frequency was previously analysed over single genomes and shown to be
approximately power-law distributed with an exponent dependent on
genome size~\cite{Huynen1998,lagomarsino2009universal}.
Figure~\ref{fig2} shows that the same distribution calculated over
thousands of prokaryotic genomes has a double-scaling, with an
exponential-like decay for low ranks in its rank plot.
We tested that the shape of these component frequency distributions do
not strongly depend on the specific size or number of realizations
analyzed.  The rank plots in Figure~\ref{fig2} do not vary when
evaluated on different sub-samples of the whole data sets (Fig.~S2 of
the SM). This suggests that the frequency distributions evaluated
using the available finite empirical data sets estimate reliably the
global heterogeneity of the component usage in the systems.

\begin{figure}[h!]
\includegraphics[width=0.4\textwidth]{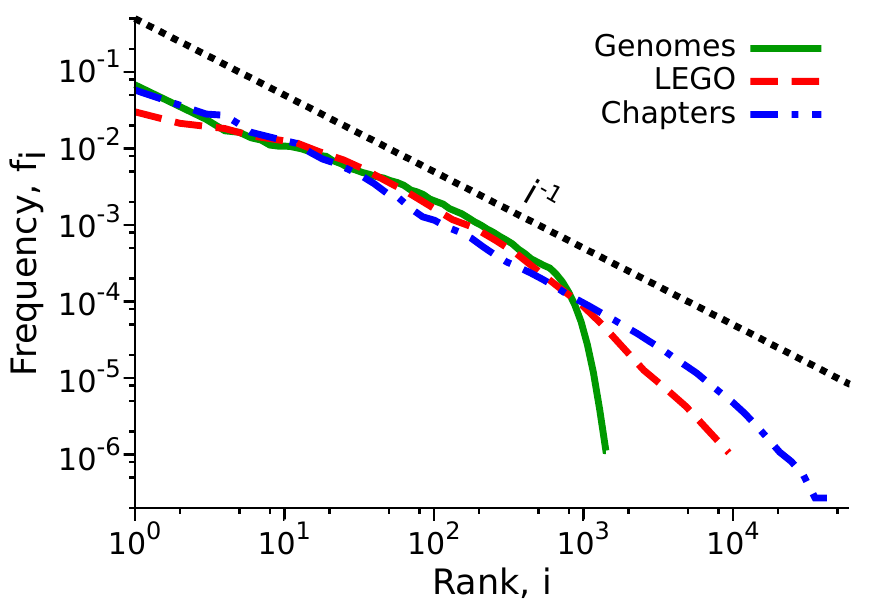}
\caption{\textbf{Different empirical component systems show similar
    component frequency distributions.}  The rank plot of the
  component frequencies is reported for the three datasets (book
  chapters, genomes, LEGO sets).  The frequency of a component is
  defined as the abundance of that component in the whole dataset
  normalized by the total number of components (Figure~\ref{fig1}).
  The three curves follow similar behavior, which can be described
  qualitatively as a power-law-like decay with exponent close to $1$
  for low ranks (high frequency), and a faster dataset-specific decay
  for higher ranks. }
\label{fig2}
\end{figure} 

We aim to evaluate also the distribution of shared components,
$\lbrace o_i \rbrace$, and how much of its features can be explained
from other measurable quantities, namely, the component frequencies,
the realization sizes $\lbrace M_j \rbrace$ and the number of
different components in the universe $N$.
Figure~\ref{fig3} shows this distribution for the three data sets
considered here.  For small occurrences, the plots are compatible with
a power-law decay, with a dataset-specific  exponent.  Only for
genomes this curve is clearly U-shaped (see also Fig.~S3),
and shows a ``core'' of shared
components, i.e., protein domains shared by almost all the genomes,
together with a rich group of rare components.  Book chapters do not
show this marked behavior, due to the fact that the ubiquitous words
(e.g. articles, pronouns, prepositions) are much less than the
chapter-specific words.  Finally, LEGO sets display no core of shared
components, and this is probably due to the wide range of themes using
poorly overlapping brick types.

\begin{figure}[htb]
  \includegraphics[width=0.45\textwidth]{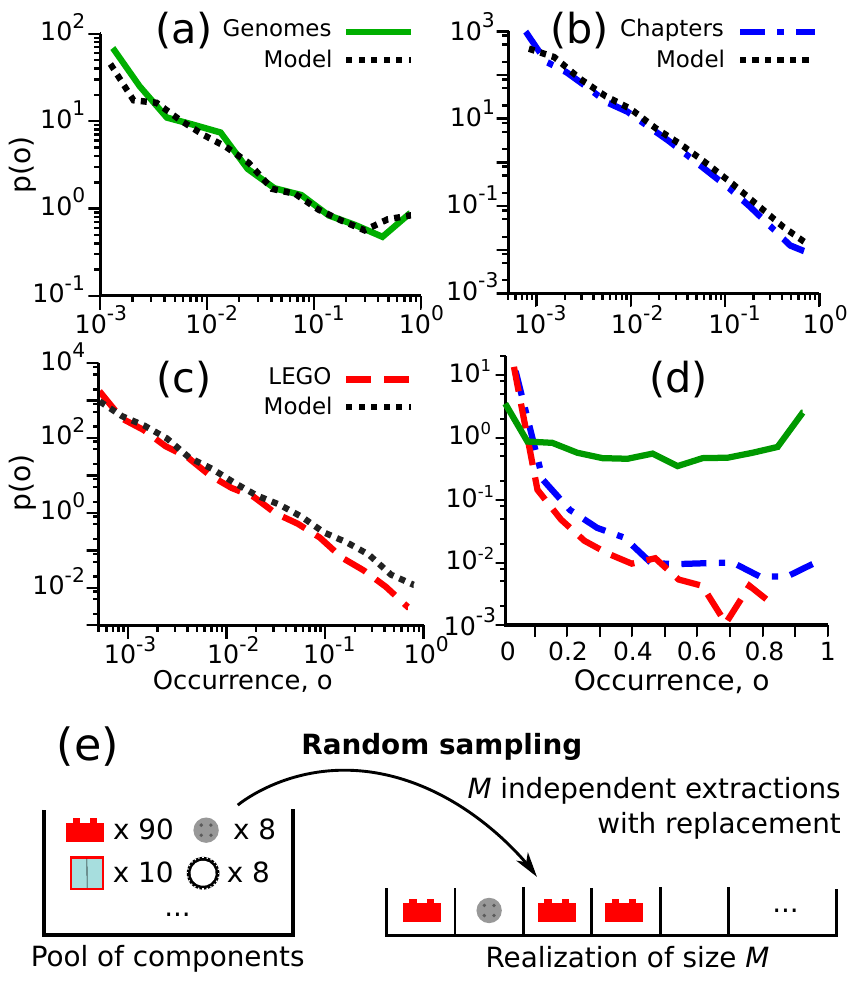}
  \caption{\textbf{The random-sampling model captures the main
      features of the empirical statistics of shared components.}  The
    plots show the normalized distribution $p(o)$ of component
    occurrences, quantifying the statistics of shared components for
    the three datasets: genomes (a) book chapters (b) and LEGO sets
    (c).  The log-log scale highlights the power-law like decay. The
    black dashed lines represent the prediction of the random-sampling
    model assuming the empirical component frequencies and realization
    sizes.  The model reproduces very well the power-law decay, but
    may differ quantitatively from the empirical laws in the
    high-occurrence region. Panel (d) plots the same quantities in
    log-lin scale, to highlight the quantitative differences between
    systems and the presence/absence of a peak of core components. 
    Note that the different range of the y-axis values with respect to 
    previous panels is  due to the different binning procedures,
    logarithmic vs linear.  
    (e) Scheme of the random-sampling process: samples of size $M$ are
    generated from independent draws from the ``universe'' of all
    possible components with their specific abundances. Therefore, the
    probability of a component extraction is proportional to its
    global abundance, i.e., the sum of its abundances over all
    realizations of the systems.}
\label{fig3}
\end{figure}

\subsection{A random-sampling model as a minimal model for component systems with defined component frequencies}
\label{sec:random_sampling}

In order to identify the statistical consequences of a heterogeneous
usage of components on the statistics of shared components, a suitable
model is needed.  In particular, we would like to generate system
realizations starting from a fixed component frequency distribution
without any additional functional information or constraint.  To this
end, we employ a random-sampling
procedure~\cite{Eliazar2011,vanLeijenhorst2005,Petersen2012,Gerlach2013}
that builds artificial realizations through an iterative random
extraction (with replacement) of components from their frequencies
$\lbrace f_i \rbrace$ in the whole system.  Each realization size $M$
is specified by the number of random extractions.

More precisely, the following prescriptions (Fig.~\ref{fig3}e)
define the random-sampling model that will be used in the following.
(i) The component abundance rank distribution is assumed to be a
universal property of the component system and well represented by the
empirical overall abundances (see Fig.~S2 and the SM for a discussion
of this assumption).  (ii) The extraction probability of a component
is proportional to its overall abundance.  (iii) A realization of size
$M$ is generated by $M$ independent extractions from the pool of
components.
Statements (ii) and (iii) define a multinomial process.  Given a
normalized list of component frequencies $\lbrace f_i \rbrace$, $i =
1, \ldots N$ (where N is the size of the available ``vocabulary''),
and the size $M$ of the realization, the probability of a specific
configuration $\lbrace n_{1}, n_{2}, \ldots, n_{N} \rbrace$, where
$n_i$ is the number of the components with frequency $f_i$ is
\begin{equation} \label{eq:multinomial}
 P(n_{1}, n_{2}, \ldots, n_{N};M) = \frac{M!}{\prod_{i=1}^N n_{i}!} 
 \prod_{i=1}^N f_i^{n_{i}}
\end{equation}
under the constraint that $\sum_{i=1}^N n_i = M$.  Note that the
expected value of $n_i$ is $M f_i$. Therefore, on average the global
abundance distribution is conserved in each realization.  In other
words, the component composition in each realization is a sampled copy
of the universe, without any of the possible complex correlations
which may follow from architectural and functional properties of an
empirical system.  

For example, in the context of bacterial genome evolution, the
random-sampling model translates into a scenario in which there is
continuous and completely random horizontal gene transfer (exchange of
genetic material) between species~\cite{lobkovsky2014}.  Thus, genome
composition would simply reflect the pan-genome abundances of protein
domains.  While horizontal gene transfer is indeed a major force in
bacterial evolution~\cite{Koonin2008,Soucy2015,Dixit2015}, several
additional genome-specific functional constraints are clearly in place
in evolution~\cite{Nimwegen2003,Molina2009,Maslov2009,Grilli2011,Soucy2015}
and these are neglected by the model. Therefore, the random sampling
can be considered as a null model useful to disentangle the
consequences of the observed global heterogeneity in the component
usage from actual hallmarks of more complex functional constraints.

\subsection{The distribution of shared components is mainly a
  consequence of component frequencies, number of available components
  and realization sizes}

The fact that the distribution of shared components is qualitatively
very similar in systems that are so different triggers the question of
whether it may be an emergent statistical consequence of other system
properties.  In particular, we asked to what extent the
statistics of shared components could be a direct consequence of
component frequencies.  As explained above, this question can be
addressed quantitatively using a random-sampling model that generates
an artificial copy of the empirical system by drawing realizations
(whose sizes are fixed by the empirical ones) from the component
frequency distribution.
Figure~\ref{fig3} compares the empirical occurrence distributions with
simulations of a random sampling. The null-model curves (dashed lines)
provide very good approximations of the empirical laws, particularly
for low component occurrences. Additionally, the model matches well
the power law decay with the system-specific exponent.  Finally, the
model predicts also the qualitative behavior of core components, and
specifically that only genomes show a clear U-shaped distribution of shared
components. The relative core sizes of the three systems are also well
approximated, although there are some quantitative deviations from the
empirical values that will be addressed in detail in
Section~\ref{sec:deviations}.
These results suggest that the shape of the distribution of shared
components in the three widely different empirical systems considered
here is well described by a random-sampling model that only conserves
the empirical component frequencies, the vocabulary (i.e., the set of
possible components) and the realization sizes. The next section
provides an analytical understanding of this observation.

\subsection{A wide range of component frequency patterns lead to
  occurrence distributions with power-law decay and U-shape.}

Thus far, we have used the model only to address 
the specific statistics of component sharing of 
the empirical systems under consideration. To this end, we have
simulated the random-sampling model fixing the component
frequencies and realization sizes as in the empirical cases.
More in general, one can ask whether a power-law decaying and/or
U-shaped distribution of component occurrences are expected for a given 
distribution of component frequencies. 
To address this question, we have computed analytically the
distribution of shared components under general prescriptions for the
component frequency distributions within the random-sampling model.  

For the sampling procedure explained in
Section~\ref{sec:random_sampling}, the probability $q_i$ that a
component of rank $i$ is present in a realization of size $M_j$ is $
q_i(M_j)= 1 - (1-f_i)^{M_j}$, where $ f_i$ is the component
probability of extraction.  Therefore, the expectation value for the
occurrence of component $i$ over a set of $R$ realizations is
\begin{equation} o_i = \frac{1}{R}
 \sum_{j=1}^{R} q_i(M_j) = 1 - \frac{1}{R} \sum_{j=1}^{R}
 (1-f_i)^{M_j} \ .
\label{eq:occurrence}
\end{equation}

In order to obtain the probability distribution associated to this
rank representation, one can use the fact that the rank of a
component with occurrence $o$ is the number of components with
occurrence higher than $o$. In fact, 
these naturally correspond to components with higher frequency and thus lower rank. 
Therefore, we can write the rank $i(o)$ as  
\begin{equation} \label{eq:rank to prob}
i(o)= \text{rank}(o) = \sum_{o'=o}^{o_1} N p(o')
 \simeq N \int_o^{o_1} p(o') d o' \ ,
\end{equation}
where $o_1$ is the highest possible occurrence, which corresponds to
the component of rank 1. The function $i(o)$  is simply 
the inverse function of Eq.~\ref{eq:occurrence}. 
From the approximate integral representation of $i(o)$, the occurrence
probability distribution $p(o)$ is defined by the simple relation $\frac{d i(o)}{do} =-N p(o)$. 

Eq.~\ref{eq:rank to prob} provides a general relation between the
representation of the frequency distribution as a rank plot and the
representation as a probability distribution.  Indeed, the arguments
presented here to introduce Eqs.~\ref{eq:occurrence}
and~\ref{eq:rank to prob} have been previously used to establish the
connection between Zipf's law as a rank plot and Zipf's law as a
frequency distribution~\cite{Mitzenmacher2004}. 
\\

\subsubsection{Observed versus possible vocabulary of components
    and Heaps' law}
    
When a set of $R$ realizations of size $M$
is generated through a random-sampling procedure from a pool of
$\tilde{N}$ possible different components with their probabilities
of extraction $\{f_i\}$, the expected size $N$ of the vocabulary
that is actually sampled can be expressed
as~\cite{vanLeijenhorst2005}
\begin{equation} \label{eq:dictionary}
  N = \tilde{N} - \sum_{i=1}^{\tilde{N}} \left( 1 -f_i \right)^{MR} \ .  
\end{equation}
Thus, in general, $N\leq\tilde{N}$.

If the system size, defined by the total number of extractions $M R$,
is large enough, essentially all possible components are expected to
be sampled at least once, thus leading to the simplification
$N\simeq\tilde{N}$ that we implicitely assumed in
Eq.~\ref{eq:multinomial}.  However, in general, the observed
vocabulary in an ensemble of realizations is an increasing function of
the system size, i.e., $N(MR)$. This functional dependence is the
analogous of Heaps' law, which is the empirical power-law growth of
the number distinct components with the system size observed in
linguistics~\cite{Altmann2016,Gerlach2013}, and in
genomics~\cite{lagomarsino2009universal}.  This distinction between
the observed and the possible vocabulary of components is discussed in
more detail in the SM
and will be relevant in the following sections.
\\

\subsubsection{Analytical distribution of shared components for component frequencies with a power-law or an exponential distribution}
 
\begin{figure}[htb]
  \includegraphics[width=0.45\textwidth]{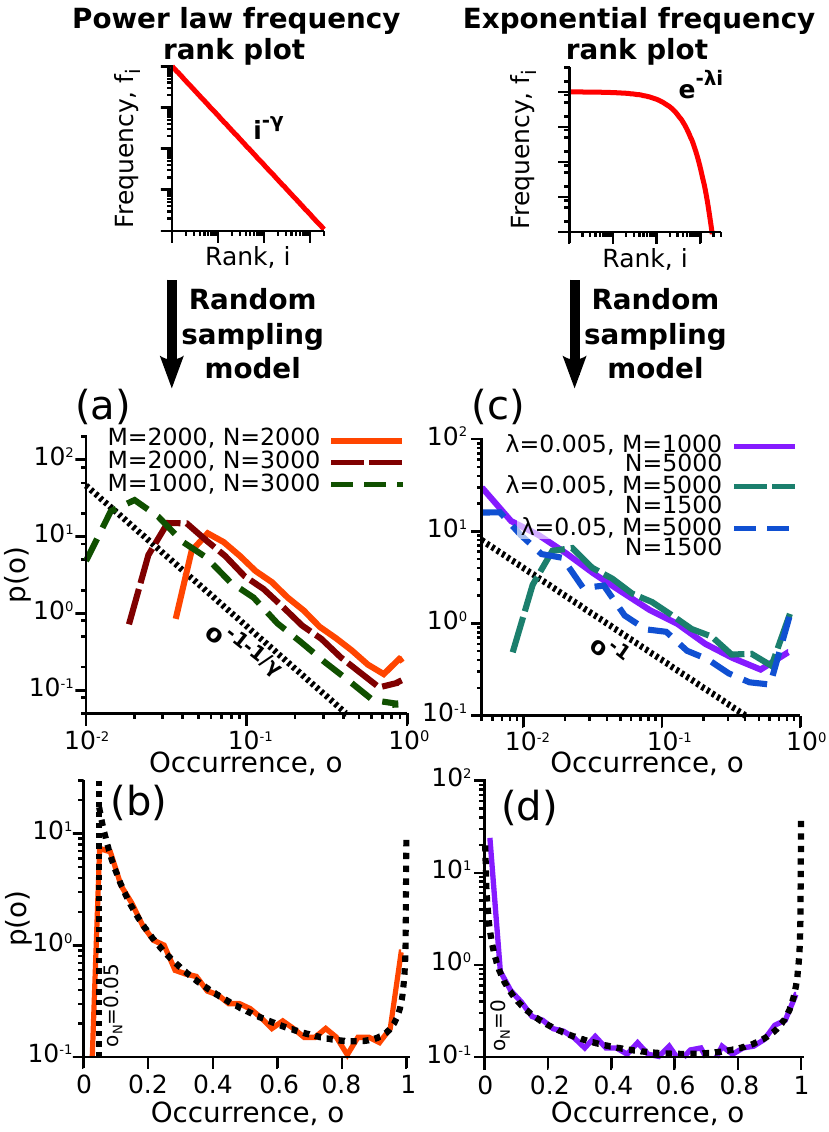}
  \caption{\textbf{Power-law decaying and U-shaped component
    occurrence distributions may descend from both power-law and
    exponential distributed universe component frequencies.} (a): A
    power-law rank-plot for the frequency (and thus for the
    abundance), whose exponent is $-\gamma$ ($\gamma=1.2$ in the
    plot), produces a power-law decay of the component occurrence
    distribution with exponent $-1-\frac{1}{\gamma}$, independently of
    the realization size $M$ and the number of components $N$ (for
    sufficiently large values of these parameters).  (b): Agreement
    between the theoretical prediction of Eq~\ref{eq:plaw_u} (black
    line) and a simulated random sampling with parameters $R =
    1000$, $N = 2000$, $\gamma = 1.2$, $M = 2000$ (the black vertical
    dashed line is the left boundary of the $p(o)$ domain).  Panels
    (c) and (d) are the counterpart of (a) and (b) for an exponential
    frequency rank plot.  In this case $p(o)$ always decreases with
    exponent $-1$, for every value of $\lambda$, $M$, and $N$
    (sufficiently large). Parameter values: $R = 1000$, $N = 2000$,
    $\lambda = 0.005$, $M = 5000$.
    Given the system sizes $MR$ in these examples,  
    the number of possible different components essentially coincides 
    with the vocabulary actually sampled , i.e., $\tilde{N}\simeq N$. } 
\label{fig4}
\end{figure}

Explicit expressions for the occurrence distribution can be derived
assuming a simple scenario, in which all realizations have the same
size $M$, and the component frequency statistics follows a prescribed
function.  We first consider the empirically relevant case of a
power-law frequency rank plot (Fig.~\ref{fig4}, left panel) defined by
\begin{equation} \label{eq:power law}
 f_i = \frac{1}{\alpha} i^{-\gamma}, \hspace{1cm}
 \alpha = 
 \sum_{i=1}^{\tilde{N}} i^{-\gamma} \ . 
\end{equation}
Under these assumptions and using Eqs.~\ref{eq:occurrence}
and~\ref{eq:rank to prob},  the exact expression of the
occurrence distribution can be calculated:
\begin{equation} \label{eq:plaw_u}
 p(o) = \frac{ \left( 1-o \right)^{\frac{1}{M}-1}}{\gamma M N 
 \alpha^{\frac{1}{\gamma}} \left( 1 - (1 - o)^{\frac{1}{M}} 
 \right)^{\frac{1}{\gamma}+1}}  \ .
\end{equation}
The distribution is defined in the interval of occurrences $[o_{N};o_1]$, 
where $o_i$ is computed by Eq.~\ref{eq:occurrence} and $N$ is the effective 
or observed component vocabulary, which can be a function of the 
system size, i.e., $N(MR)$, as described by Eq.~\ref{eq:dictionary}.
Considering the limit of small occurrences and large sizes, i.e., $o \ll 1$ and 
$M \gg 1$, one finds precisely the empirically observed  power-law decay. 
Specifically, in this limit the occurrence distribution takes the form 
\begin{equation} \label{eq:plaw_u_lim}
 p(o) \simeq \frac{M^{\frac{1}{\gamma}}}{\alpha^{\frac{1}{\gamma}} 
 \gamma N} o^{-\frac{1}{\gamma}-1} \ ,
\end{equation}
where the power-law exponent depends only on the exponent $\gamma$ of
the frequency rank-plot .

Analogous calculations (details in the SM) can be performed assuming a
frequency distribution described by an exponential rank plot $f_i \sim
e^{-\lambda i}$ (right panel of Figure~\ref{fig4}).  In this case, the
distribution of shared components, for large enough realizations
$M\gg1$, has the expression
\begin{equation} \label{eq:occ_exp}
p(o) \simeq \frac{(1-o)^{-1}}{N \lambda \log{\left[ (1-o)^{-1} \right]}}. 
\end{equation}
Interestingly, for rare families the above expression further
simplifies to a power-law decay
\begin{equation} \label{eq:occ_exp2}
p(o) \simeq \frac{1}{N \lambda} o^{-1},
\end{equation}
with a ``universal'' exponent $-1$.  This indicates that also systems
with a heterogeneous but more compact frequency distribution are
expected to show a power-law decay in the occurrence distribution.
Figure~\ref{fig4} shows the agreement between these predictions and
simulations of the random-sampling model for the two illustrative
examples of a power law and of an exponential distribution of
component frequencies. 
These analytical predictions have a dependence on the sampled vocabulary
$N$ and are expected to hold even if this is actually smaller 
than the total number of possible components $\tilde{N}$ (Fig.~S5). 
The effects of a dependence of the observed dictionary on system size (i.e., Heaps' law $N(MR)$) 
become relevant and has to be taken into account when comparing statistical features of ensembles of 
realizations with different sizes $MR$.
\\

\subsubsection{Shape of the distribution of shared components and 
  rescaling properties}

We now turn our attention to the conditions for a U-shaped
distribution of shared components in the random-sampling model.
Figure~\ref{fig4}ac already show that the decay of the occurrence of 
rare components is only set by the exponent $\gamma$ as described by 
Eq.~\ref{eq:plaw_u_lim}, but for different values of $M$ and 
$N$ the distribution may or may not display a significant 
fraction of core components.
Additionally, Figure~\ref{fig4}bd proves that
equations~\eqref{eq:plaw_u} and~\eqref{eq:occ_exp} can capture
quantitatively the occurrence distributions and thus can well describe
the relative proportion of core and specific components.
In order to understand under what conditions this distribution becomes
clearly U-shaped for an underlying power-law frequency distribution, it is
useful to note a rescaling property of Eq.~\ref{eq:plaw_u}.  Taking
the limit of large realizations $M \gg 1$, Eq.~\ref{eq:plaw_u} becomes
\begin{equation} 
\label{eq:plaw_u_largeM} 
  p(o) = k(\gamma,M,N) 
  \frac{(1-o)^{-1}}{\gamma
  \left(-\log(1-o)\right)^{1+\frac{1}{\gamma}}} ,
\end{equation}     
which depends only on two parameters, $\gamma$ and the rescaling parameter
\begin{equation} 
  \label{eq:k} 
  k(\gamma,M,N) = \frac{M^{\frac{1}{\gamma}}} 
  {\alpha^{\frac{1}{\gamma}} N} \  .
\end{equation}
This rescaling property shows that the statistics of component sharing
is actually a function of a specific combination of realization sizes
(e.g., text lengths) and of the range of possible components
(e.g., the observed vocabulary). 
Specifically, the functional form of the distribution is purely
defined by the exponent $\gamma$, while the rescaling parameter $k$
sets the normalization factor and the range of possible occurrences.
In fact, the analytical expression of the occurence corresponding to
the distribution minimum, i.e.,
$o_{min} = 1 - e^{-1-\frac{1}{\gamma}}$, is only a function of
$\gamma$, while the minimum possible occurrence value
$ o_N \simeq 1 - e^{-k^\gamma}$ scales with $k$.  Therefore, a
U-shaped occurrence distribution should be generally expected for
component systems with highly heterogeneous component frequencies
since the power-law decay and the presence of a minimum before the
core are robust features with respect to system parameters.  This is
confirmed by the analysis of component systems with different values
of $k$ and $\gamma$ (illustrative examples in Figure S7): the system
specificities set the power-law decay of the left part of the
distribution, its support, and the relative proportion of core and
rare components, but the U-shape is conserved.  However, this shape
can be more or less symmetric and more or less clearly evident
depending on the actual size of the core fraction. The following
section discusses in detail the non-trivial dependences of the core
size on system parameters.

For the case of component frequency distributions with an exponential
rank plot, the statistics of shared components (Eq.~\ref{eq:occ_exp})
is a function of a single effective parameter $\lambda N$, and does
not depend on the realization sizes $M$.  In other words, the shape of
the distribution, and whether it is clearly U-shaped, only depend on
the decay of component frequencies and on the total number of
components.  In fact, occurrence distributions corresponding to
different exponential frequency rank plots collapse if $\lambda N$ is
constant, even if the realizations have widely different size. This is
shown in Figure S4 of the SM. 
\\

\subsubsection{The core size}
\label{core_par}

We can  estimate the ``core size'' by computing the fraction of 
components with occurrence greater than a given arbitrary occurrence 
threshold $\theta_c$ as a function of the only two effective 
parameters $\gamma$ and $k$.  
Integrating Eq.~\ref{eq:plaw_u} between $\theta_c$ and the 
maximum occurrence $o_1$, and then taking the limit $M\gg 1$, 
this quantity reads
\begin{equation} 
  \label{eq:common}
 \begin{cases}
  c = 1   & \text{if} \hspace{0.3cm} o_{N} \geq \theta_c      \\
  c = k \left[ -\log(1 - \theta_c) \right]^{-\frac{1}{\gamma}} & \text{otherwise}
\end{cases} \ ,
\end{equation}
where $o_N$ is the left boundary of the occurrence distribution, 
corresponding to the component with lowest frequency. 

\begin{figure}
  \includegraphics[width=0.48\textwidth]{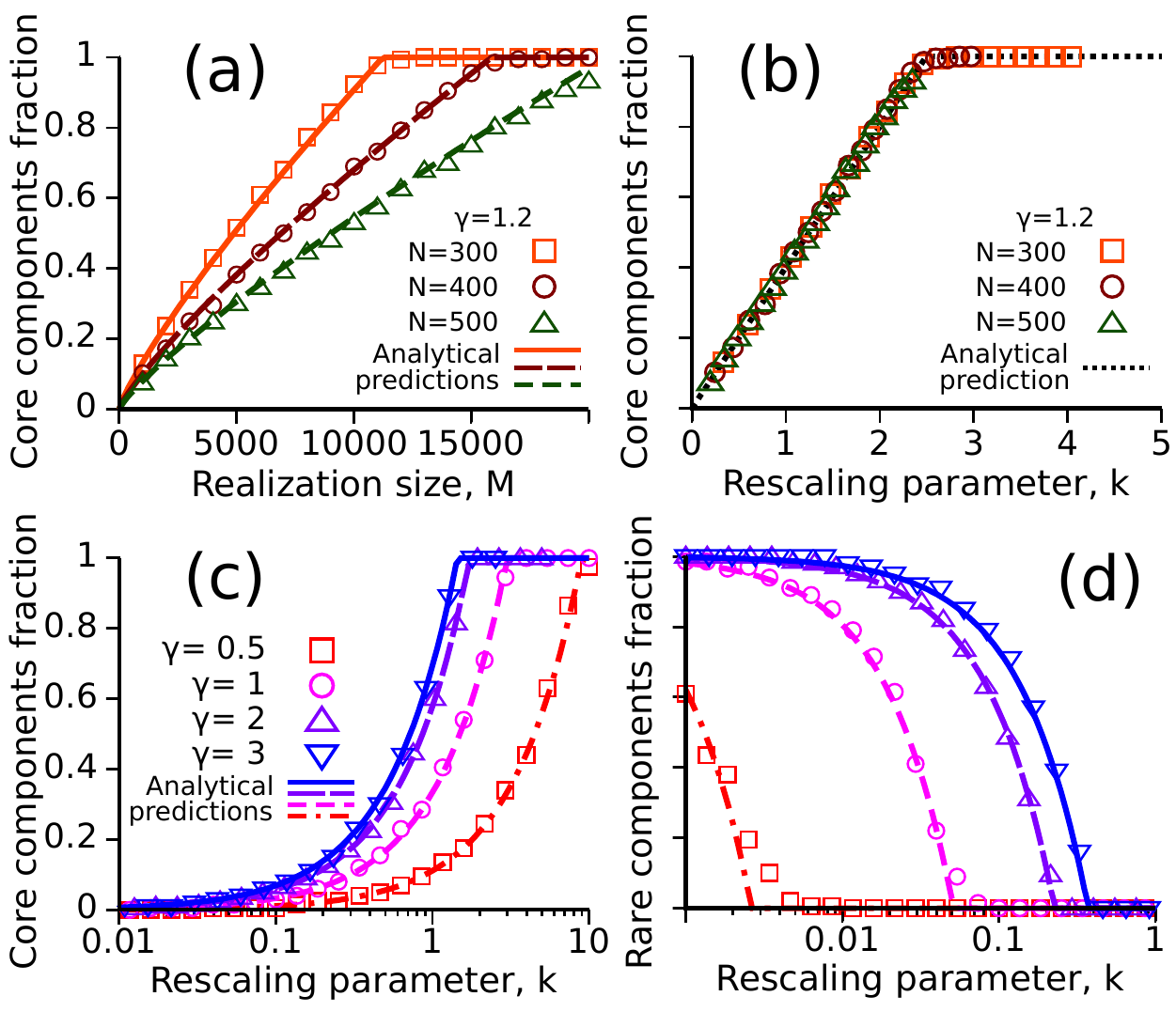}
  \caption{\textbf{Scaling of the distribution of shared components
      and fraction of rare and core components.}  (a) The fraction of
    core components (defined by the occurrence threshold $ o> \theta_c
    = 0.95$) for a power-law component frequency distribution with
    exponent $\gamma = 1.2$, plotted as a function of component size
    $M$ for three values of realization number $N$.  (b) Collapse of
    the curves shown in panel (a) when plotted as a function of the
    rescaled parameter $k$, defined in Eq.~\ref{eq:k}.  (c) and (d):
    fraction of core and rare ($o <  0.05$) plotted as a
    function of $k$ for different values of $\gamma$.  For
    sufficiently large $k$ (i.e. typically when $M$ dominates over
    $N$), the fraction of core components saturates to $1$.
    Conversely, the fraction rare components drops to zero for
    increasing $k$.  Symbols refer to numerical simulations of the
    random-sampling model, while the lines are the theoretical
    predictions of Eq.~\ref{eq:common}.}
\label{fig5}
\end{figure}

Starting from this estimate of the core size, 
Figure~\ref{fig5}ab show how the scaling property is verified in
simulations.
Fig.~\ref{fig5}c compares the analytical predictions for the core size
with simulations for different values of $\gamma$, 
showing perfect agreement.  Equally, one can obtain
analytical estimates for the fraction of rare components (occurrence
below a fixed threshold), which are tested in Fig.~\ref{fig5}d.
Thus, with increasing $k$, core families increase linearly with a
$\gamma$-dependent slope until all components are shared, and
concurrently rare components decrease linearly until they hit zero
(when the lower cutoff of occurrence exceeds the chosen threshold
value).
Component number and realization size only enter through the
combination defined by the rescaling parameter $k$. This phenomenology
fully characterizes the distribution of shared components with varying
parameters. 

The general relation (Eq.~\ref{eq:common}) between the core size
and the rescaling parameter $k$ translates into different
dependences of the core size on the typical realization size $M$,
depending on the relation between the system size $MR$ and the total
number of accessible components $\tilde{N}$.
While this issue is discussed in more detail in the SM, it is
easy to intuitively understand the different regimes.  
For large enough systems, all possible components $\tilde{N}$ are 
expected to be sampled at least once, thus making the observed vocabulary
$N\simeq \tilde{N}$ a constant parameter. This is the regime
considered in Fig.~\ref{fig5}a.  In this regime, Eq. \ref{eq:common}
simplifies to the simple scaling $c \sim \frac{M^{\frac{1}{\gamma}}}
{\tilde{N}}$.
On the other hand, in several empirical systems the observed
vocabulary is a function of the system size, and typically with the
power-law dependence $N(MR) \sim (MR)^\beta$ (with $\beta < 1$)
called Heaps' law.
Thus, in general, the core fraction is expected to show the more
complex dependences $c\sim M^{\frac{1}{\gamma} -\beta} R^{-\beta}$.
However, a random-sampling procedure starting from a Zipf's law
described by Eq.~\ref{eq:power law} leads to the approximate
relation $\beta \simeq 1/\gamma$ between the exponents of Zipf's and
Heaps' law~\cite{Eliazar2011,vanLeijenhorst2005, lu2010zipf}.
Therefore, in this regime the core fraction becomes only a function
of the number of realizations as $c\sim R^{\frac{1}{\gamma}}$.
These different scaling relations in different regimes are tested in
Figure S6.
Note that the absolute number of core components $c N$, as estimated
from Eqs.~\ref{eq:k} and~\ref{eq:common}, is instead always
independent from the number of realizations, even in the regime
where Heaps' law is expected to hold (Figure S6).  

For component frequency distributions with an exponential rank
plot, the sampling procedure leads to an occurrence distribution
that is independent from the realization size $M$ (Eq.~\ref{eq:occ_exp}).  
However, the exact analytical prediction for the core size (the analogous 
of Eq.~\ref{eq:common}) still has a dependence on $M$.  
But this is due to the residual dependence of the maximum occurrence 
values ($o_1$) on $M$ and does not affect the shape of the distribution. 
This last technical point is discussed in more detail in the SM.

\subsection{Empirical distributions of shared components satisfy the
  relations predicted by the random sampling.}

\begin{figure}
\includegraphics[width=0.45\textwidth]{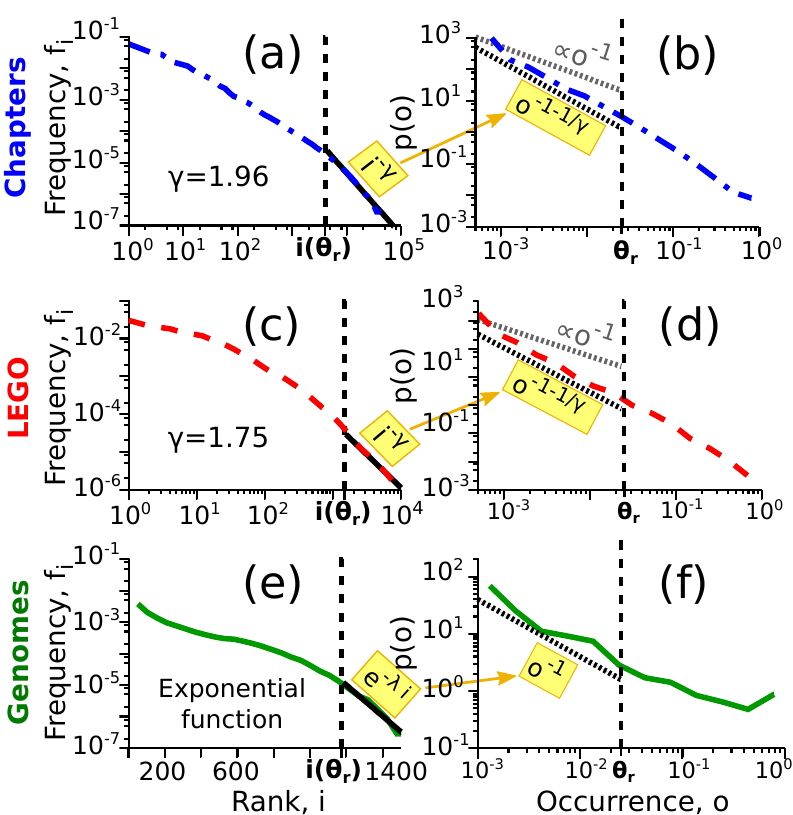}
\caption{\textbf{The relation between the exponents of frequency rank
  plot and occurrence distribution is satisfied in all the three
  datasets.}  The plots consider the low occurrence region, below
  the arbitrary threshold $\theta_r = 0.025$ which corresponds to the
  high-rank region above $i(\theta_r)$ in the frequency rank plot (see
  main text).  Panel (a) and (b) refer to book chapters, for which the
  tail of rank plot is a power law with exponent $\gamma = 1.96$,
  which implies a power law decay of $p(o)$ with exponent
  $1+\frac{1}{\gamma} = 1.51$.  Panel (c) and (d) show the LEGO
  dataset ($\gamma = 2.8$, $1+\frac{1}{\gamma} = 1.36$).  Panel (e)
  and (f) correspond to protein domains in genomes, where the best fit
  of the tail region the rank plot is an exponential function (note
  that (e) is in linear-logarithmic scale), which implies a power law
  decay with exponent $-1$. }
\label{fig6}
\end{figure}

One can ask whether the general analytical predictions discussed in
the previous section can be applied to empirical data.
In particular, we first asked how the power-law decay exponent of the
distribution of shared components relates to the component
frequency rank plot in empirical systems,
and if this relation follows our analytical prediction.  
An analytical mapping would give a more synthetic 
and powerful description than the direct simulations discussed in
Fig.~\ref{fig3}. 
Importantly, the analytical formulas for the distribution of shared
components are derived under the hypothesis of a pure power-law 
or exponential component frequency rank plot. 
However, the three empirical datasets (as previously discussed), 
show  a double-scaling frequency distribution. 
To override this issue, we
restricted the frequency rank plot range in which the predictions are
applicable.  The procedure to perform this comparison is described in
Fig.~\ref{fig6}. 

First, we chose an arbitrary threshold $\theta_r$ defining the rare
components and we mapped it to the frequency rank plot (assuming the
model), by using the inverse function of Eq.~\ref{eq:occurrence}. The
frequency rank associated to the occurrence threshold $\theta_r$,
$i(\theta_r)$ in the figure, is the rank above which the model
prediction for the decay of the distribution of shared  components 
should apply as long as $i(\theta_r)$ does not cross the position of 
the change in scaling. 
In other words, since in the model there is a monotonic relation 
between occurrence and frequency (Eq.~\ref{eq:occurrence}), all
 components with rank greater than $i(\theta_r)$ (and frequency
smaller that $f_{i(\theta_r)}$) are assumed to be the components with
occurrence lower than $\theta_r$.
We then estimated the behavior of the frequency rank plot in the
high-rank region (after $i(\theta_r)$) as the best fit with a
power-law function or an exponential. This leads to a prediction for
the decay exponent of the distribution of shared components (using
Eq.~\ref{eq:plaw_u_lim} or Eq.~\ref{eq:occ_exp2} for the exponential
case) in the range $[o_N, \theta_r]$.
Fig.~\ref{fig6} shows that the predicted decay exponents correspond
well with the data. 

The random-sampling model also gives qualitative analytical
predictions for the expected fraction of core components, and thus for
the expected shape of the distribution of shared components for a
given empirical system.  While the analytical relations between
exponents applied in Figure~\ref{fig6} do not depend on the
realization sizes, the analytical formulas for the fraction of core
components (see e.g. Eq.~\ref{eq:common}) were derived assuming
realizations of fixed size $M$.  The actual size distributions for the
three empirical systems are quite broad (Figure S1), but we can still
use the analytical framework to get an estimate of the core fraction
considering the average realization size of each empirical system.
Following the same line of reasoning as for the low-occurrence tail of
the distribution of shared components, we can use a restricted region
of the frequency rank plot.  In this case, the low-rank region (with
exponent around $1$ for all the datasets, see Fig.~\ref{fig2}) is
expected to contain the core components.
Therefore, the parameter $\gamma$ can be fixed to $1$, implying that
the fraction of core components, given by Eq.~\ref{eq:common}, should
be simply proportional to the rescaling parameter $k$
(Eq.~\ref{eq:k}). 
However, the  normalization factor $\alpha$, which is
present in the definition of $k$ and defined in Eq.~\ref{eq:power
  law}, takes an approximately constant value with respect to $\tilde{N}$ for
large values of $\tilde{N}$, as it is the case for the empirical examples
considered.  As a consequence, the core fraction should be simply
proportional to $\frac{M}{N}$.  This estimate can be used to explain
why the core fraction is much larger in genomes than in the other two
empirical systems (see Figure~\ref{fig2}d).  In fact, genome sizes are
typically of the same order as the total number of families ($M \simeq
3000$, $N = 1531$, see Figure S1) leading to a large expected core. By
comparison, book chapters have similar realization sizes but a much
larger vocabulary ($N \simeq 50000$), and LEGO sets have very small
sizes ($M \simeq 100$) compared to vocabulary size ($N \simeq 13000$).

More in general, Eqs.~\eqref{eq:k} and~\eqref{eq:common} lead to a
scaling estimate (dependent on the decay of the frequency rank plot)
as a function of the system parameters $M$ and $N$, which can be
applied to data, in order to generate expectations for the core
components.  For example, for Zipf-like (exponent -1) frequency
distributions, we expect the absolute number of core components to be
linearly dependent on the average size of realizations $M$, and
essentially insensitive to the vocabulary size $N$ and the total
number of realizations $R$.  In genomics language, this would imply
that the number of core protein domains does not directly depend on
the number of sequenced genomes but only on their sizes and on the
total number of different protein domains discovered.  Note that
adding new genomes to the data set is not expected to alter the
power-law exponent $\gamma~\simeq1$ of the global frequency
distribution for high-frequency components, since it does not change
if the distribution is evaluated on sub-samples of the empirical
dataset
(Figure S2).\\
As previously discussed, the core fraction, instead of the
absolute number of core components, is expected to have a more
complex dependence on the typical realization size $M$ and on the
number of realizations $R$.  Moreover, in empirical systems these
relations are further complicated by the fact that the frequency
distributions cannot be described by simple power-laws (Fig.~\ref{fig2}).
Nevertheless, the relation between the core fraction and the average
realization size predicted by a random-sampling model can be tested numerically, 
as  Figure~\ref{fig7}a shows for prokaryotic genomes, and seems 
accurately verified and roughly linear in the tested range of sizes.  
However, the predicted fraction of core components is actually 
much smaller than the empirical one.  
This highlights the presence of additional functional constraints 
and/or specific correlations in the empirical system that the model 
can not capture.  
The next section addresses this point more in detail.

\subsection{Deviations from the random-sampling predictions can highlight system-specific properties}
\label{sec:deviations}

\begin{figure}
\includegraphics[width=0.48\textwidth]{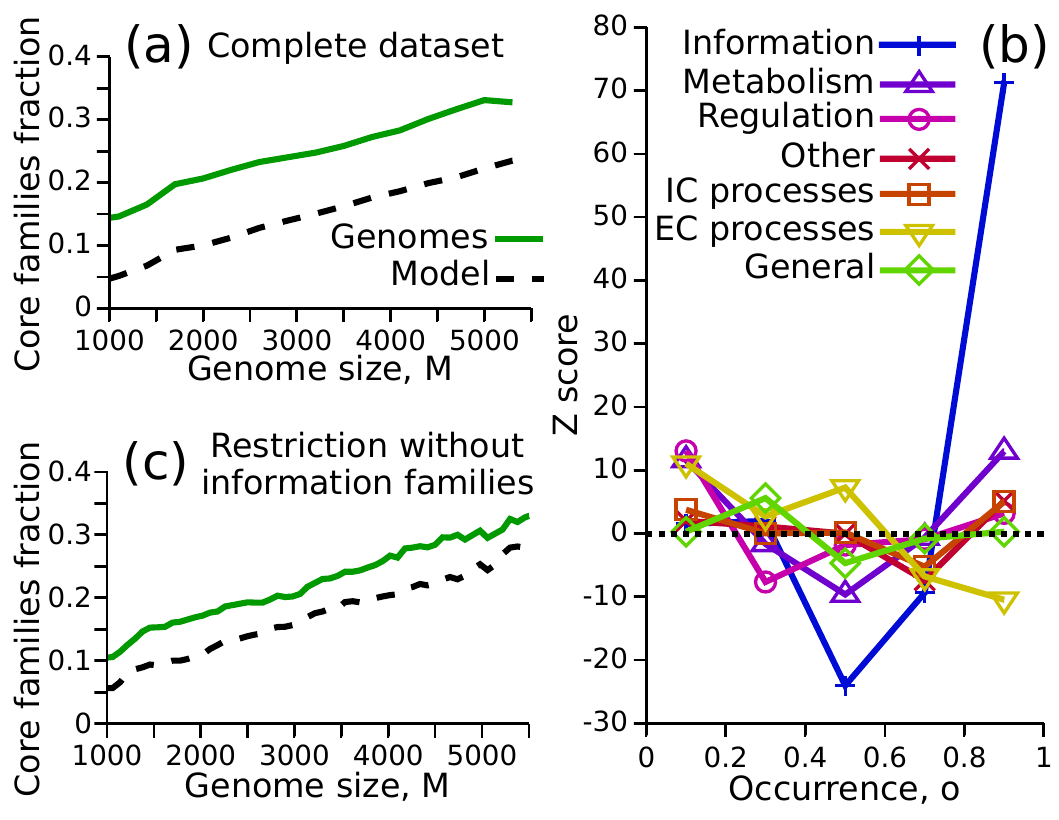}
\caption{\textbf{Specific functional constraints can be detected by
    deviations from the predictions of a random sampling.}
  a) Fraction of common protein domain families as a function of the
  genome sizes.  Each point of the curves corresponds to the core
  families ($o>\theta_c=0.95$) given the occurrence distribution of a
  genomes' subset whose sizes are inside a certain window.  The
  average of the size windows defines the x axis.  
  b) Enrichment analysis in the occurrence distribution for specific functional categories. 
  Considering  domain families relative to a single functional category, 
  their relative component occurrence distribution  was evaluated for an ensemble of 
  systems built with a random sampling.  
  From this, the average value and the standard deviation 
  for the expected fraction of components at each occurrence value $o$ can be calculated. 
   This provides a measure (Z score) of over- or under-representation of domain families 
  belonging to each functional category in the empirical dataset. 
  c) Excluding from the analysis the domain families associated to information processes 
  (i.e., DNA replication, transcription and translation) 
  significantly reduces the offset between the random-sampling prediction and 
  the empirical trend.}
\label{fig7}
\end{figure}

Beyond the striking agreement with null predictions for shared
components, the deviations from sampling can be used to quantify
specific functional and architectural features of a component system.
While the scope of this work is to highlight the common trends and
their origins, we discuss a specific example, in order to show the
feasibility of this procedure.
Of the three data sets considered here, the case where the clearest
deviations emerge are genomes.  For example, Figure~\ref{fig7}a
illustrates how the random sampling underestimates the empirical core
size by a constant offset, for genomes of increasing size.
Generally speaking, this larger core of components is due to the
components that tend to occur in most realizations, but in few
copies. The natural explanation is that there are specific basic
functions that are essential for all (or most) genomes, but the
domains involved in these functions are not necessarily needed in many
copies per genome, and thus their presence in all realizations does
not simply correlate with high global abundances as the random
sampling would entail~\cite{Koonin2003}.

To test this hypothesis, we divided the domain families in functional
categories (see Section~\ref{data} for the functional annotation), and
tested if most of the deviations from the random-sampling prediction
can be ascribed to the statistics of domains belonging to specific
categories. The result of this analysis is reported in
Figure~\ref{fig7}b.  Different parts of the distribution of shared
components are indeed enriched in components of different biological
functions with respect to the random-sampling expectation.  In
particular, protein domains that play a functional role in information
processes - such as DNA translation, DNA transcription, and DNA
replication- are clearly enriched in the core.  At the same time, they
seem statistically under-represented at occurrences around 0.6.  These
two deviations can be explained as two sides of the same coin if this
category contains domain families that empirically occur in all
genomes but in a single copy per genome.  Indeed, the global frequency
(i.e, across all genomes) of families that are both single-copy and
ubiquitous is $f= \frac{R}{R M} = 1/M$.  Therefore, their occurrence
predicted by the random-sampling model is $o=1-(1-\frac{1}{M})^M = 1-
e^{M \text{log}(1-\frac{1}{M})}\simeq 1- e^{-1}\simeq 0.6$ (where the
rough approximation holds for large enough $M$), thus naturally
leading to an excess of those families in the core and to a depletion
around $o\simeq 0.6$.

The observation of a strong presence of protein domains related to
basic cellular function in the core genome is not
new~\cite{Koonin2003,Koonin2008}.  However, the random-sampling model
allows in principle to distinguish families whose presence in the core
could be simply explained by their high abundance in the pan-genome
and thus it would be expected also in a simple scenario of random gene
exchange.  Finally, the observed correlation between biological
functions and deviations from random sampling predictions seems
coherent with a picture, recently proposed~\cite{lobkovsky2013gene},
in which natural selection and functional constraints have played an
important role in defining the empirical U-shaped distribution of gene
occurrences.

\section{Discussion and Conclusions}

This work employs a simple statistical model based on random sampling
to describe the distribution of shared components in complex component
systems. A similar approach was employed in quantitative linguistics to
explain how the dictionary used in a text scales with text size as
measured in number of words (the so-called ``Heaps'
law'') while assuming  Zipf's law for component 
frequencies~\cite{Eliazar2011,heaps1978information,vanLeijenhorst2005,
  lu2010zipf, Font2015}. We extended the model to show that there
is a general link between the heterogeneity in component frequency and
the statistics of shared components, regardless of the mechanisms that
generate heterogeneity.
Consequently, models or generative processes able to explain the
heterogeneity in component frequency implicitly carry predictions for
the statistics of shared components.

The striking similarities of laws governing both component abundance
and occurrence found in empirical systems of very different origins
(LEGO sets, genomes, book chapters) support the idea that the concept
of ``component system'' defined in this work can capture in a unified
framework a large class of complex systems with some common global
properties.  Different component systems, besides having specific
architectural constraints, may show convergent phenomena in terms of
global statistics.  Such ``universal'' phenomena may be regarded as
emergent properties due to system heterogeneity, which transcend the
specific design, generative process or selection criteria at the
origin of a system. Analogous phenomena occur, for example, in
ecosystems, where emergent species-abundance distributions appear for
forests, birds or insects~\cite{Hubbell2001}.

Beyond the examples considered here, modular systems in a wide range
of disciplines can be represented as component systems.  Developing a
common theoretical language for such systems can help the exchange of
ideas, models and data-analysis techniques between distant communities
of researchers~\cite{Holovatch2017}.  
For example, the statistics of component sharing considered here plays
a central role in
genomics~\cite{lapierre2009,lobkovsky2013gene,Koonin2011} but is
relatively unexplored in the context of natural
languages~\cite{Altmann2016}.  Conversely the random-sampling approach
used here was developed in quantitative
linguistics~\cite{Eliazar2011}, and this work shows that it is
applicable to other systems, including the detection of functional
constraints in prokaryotic genome evolution.

An important result of this work is a proof of the clear link between
the heterogeneity of component abundance in a system and the
statistics of shared components. This link is consistent with data
from three very different empirical systems and  well captured by the random-sampling model. 
The fact that emergent patterns can be explained by largely null
models resembles again the case of biodiversity, where neutral
theories ignoring species interactions and competitive exclusion
appear to capture many of the emerging trends of species
abundance~\cite{Hubbell2001,Azaele2016}.

If the trends of component sharing of generic component systems are to
be regarded as largely null and due to components heterogeneity,
system-specific investigations should be informed of this general
trend.
Quantitative null models, such as the one provided here, may be crucial
for identifying dataset-specific deviations that are related to
functional reasons or constraints.  In the data considered in this
work, the patterns of shared components show differences between
empirical data and the null model in some cases.  This is particularly
true in the genomic context, where the differences can indeed be
traced back to functional constraints in genome composition.
Therefore, the framework can be useful to pinpoint hallmarks of
functional design and distinguish them from statistical effects,
particularly for the detection of causality, dependency and
correlation structures between components from occurrence patterns.

Once a null model is defined, these features can emerge as significant
deviations from the null behavior, for example as violations of the
constraints linking different global statistics such as the abundance
rank plot, the distribution of shared components and Heaps' law.
We have considered here a specific example for the case of shared
protein domain families in genomes (Fig.~\ref{fig7}), but this
question still needs to be approached systematically.  In this
specific case, core components are particularly enriched by specific
functional classes of components with respect to the random-sampling
prediction.  In evolutionary terms, the random sampling defines a
scenario in which the pan-genome fully determines the overall
abundance of the gene families in each genome, while in empirical
bacterial genomes genome-specific functional constraints are clearly
in place~\cite{Maslov2009,Molina2009,Grilli2014}.  Deviations from
the null scenario can thus highlight the role of selection for
specific functions, supporting from a different perspective the idea
that the empirical U-shaped gene occurrence distribution is affected
by selective rather than neutral
processes~\cite{lobkovsky2013gene,collins2012testing,baumdicker2012infinitely,
haegeman2012neutral}.

\section*{Acknowledgements}

We thank Erik van Nimwegen for useful discussions.
This work is supported by the "Departments of Excellence 2018 - 2022"
Grant awarded by the Italian Ministry of Education, University and
Research (MIUR) (L. 232/2016)

\bibliography{bibliography}

\clearpage

%%% SUPPLEMENTAL MATERIAL

\pagebreak
\widetext
\begin{center}
\textbf{\large SUPPLEMENTAL MATERIAL\\  
   Statistics of  shared components in complex component systems}
\end{center}

\setcounter{figure}{0}
\setcounter{section}{0}
\setcounter{table}{0}
\setcounter{equation}{0}
\renewcommand{\figurename}{Supplementary Figure}
\renewcommand{\thefigure}{S\arabic{figure}}
\renewcommand{\thesection}{S\arabic{section}}
\renewcommand{\tablename}{Supplementary Table}
\renewcommand{\thetable}{S\arabic{table}}
\renewcommand{\theequation}{S\arabic{equation}}

\section{Books composing our linguistic corpus}

\begin{table}[h!]
 \centering
 \caption{List of the books whose chapters compose the analyzed linguistic corpus. 
 Data downloaded from the Project Gutenberg:  http://www.gutenberg.org}
 \def\arraystretch{1.1}\tabcolsep=10pt
 
 \begin{tabular}{ll}
  \textbf{Title}	&	 \textbf{Author}\\
  \hline
  Alice's adventures in wonderland	&	Lewis Carroll\\
  Anna Karenina	&	Lev Nikolayevich Tolstoy\\
  A tale of two cities	&	Charles Dickens\\
  Dracula	&	Bram Stoker\\
  Emma	&	Jane Austen\\
  Great expectations	&	Charles Dickens\\
  Les miserables	&	Victor Hugo\\
  Moby Dick	&	Herman Melville\\
  Notre-Dame de Paris		&	Victor Hugo\\
  Pride and prejudice	&	Jane Austen\\
  The adventures of Tom Sawyer	&	Mark Twain\\
  The count of Monte Cristo	&	Alexandre Dumas\\
  The man in the iron mask	&	Alexandre Dumas\\
  The picture of Dorian Gray	&	Oscar Wilde\\
  The three musketeers	&	Alexandre Dumas\\
  War and peace	&	Lev Nikolayevich Tolstoy\\
  \hline
 \end{tabular}
 \label{tab:books}
\end{table}

\section{Size distributions of the realizations in the three data sets}

\begin{figure}[h!]
 \includegraphics[scale=1.6]{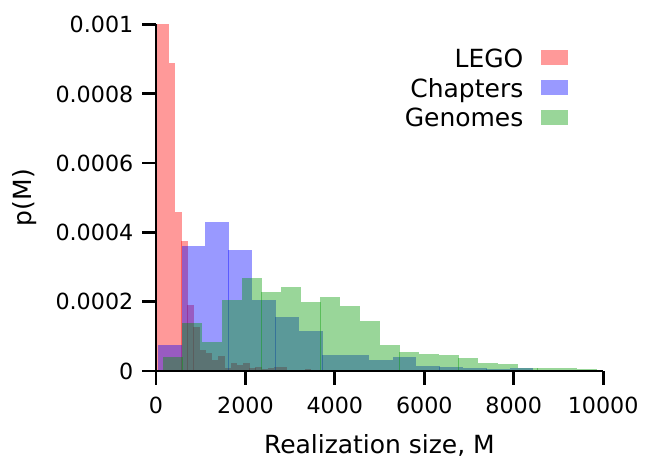}
 \caption{\textbf{Size distributions of prokaryotic genomes, book chapters and LEGO toys.}
 Probability distribution of the realization size $M$, defined as the total number of components 
 (protein domains, words or LEGO bricks) in a realization (genomes, chapters or LEGO sets).}
 \label{fig1}
\end{figure} 

\newpage
 
\section{Universality of the  component abundance distribution (Zipf's law)}

Figure \ref{fig2} tests the  component frequency rank distribution conservation when it is evaluated on different sub-sets of the total empirical data set. 
These sub-sets are composed by realizations in a fixed range of sizes, showing that the global frequency statistics does not depend on the realization sizes 
or on the number of realizations considered. 
Note that this test is necessary  to safely compare the analytical null predictions with different sub-samples of the empirical data set, as for example in Figure 7 of the main text.
Moreover, the fact that the Zipf's laws of the under-sampled datasets is essentially identical to the global one (especially in the high-frequency regime)
suggests that the observed Zipf's law is not under-sampled and can thus be considered a good estimation of the ``universal'' one.

\begin{figure}[h]
 \includegraphics[width=1\textwidth]{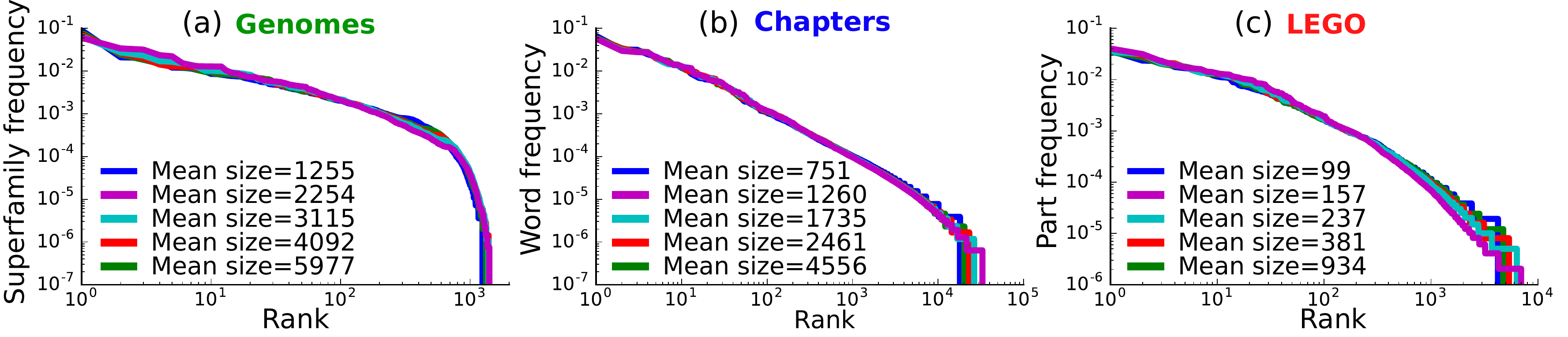}
 \caption{\textbf{Zipf's law for different sub-samples of the empirical data sets, corresponding to different bins of realization sizes.} 
  Zipf's law in genomes (a), book chapters (b), and LEGO sets (c) for several sub-sets of the data set. 
  The average sizes of the realizations in each sub-set are reported in the legend. 
  %The realizations are subdivided into groups of different sizes conserving the number of samples in each groups.
  }
 \label{fig2}
\end{figure}

\section{Robustness of the U-shaped distribution of shared components for bacterial genomes to the binning procedure.}

\begin{figure}[h!]
\includegraphics[width=0.5\textwidth]{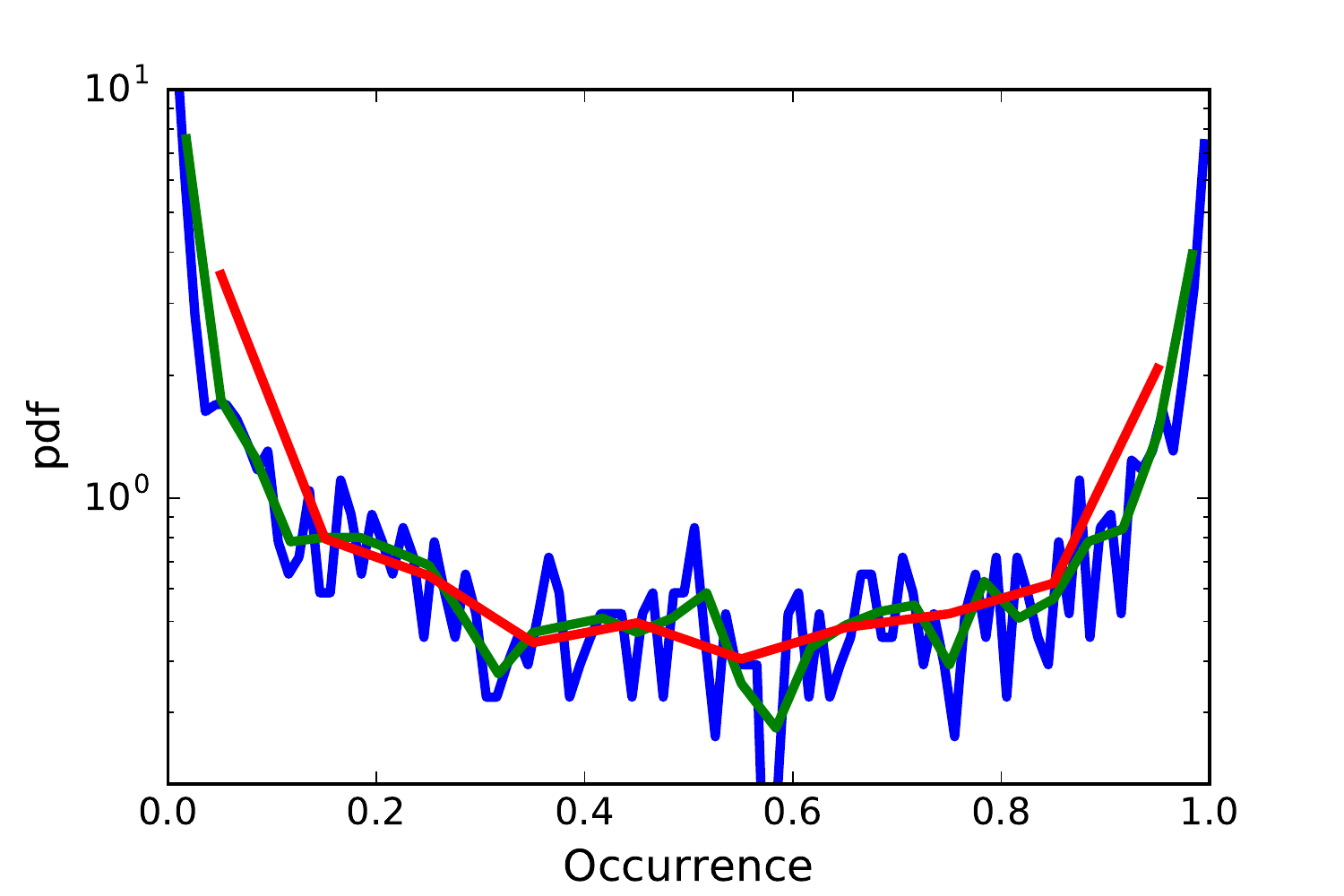}
\caption{\textbf{Robustness  of the U-shaped distribution of shared components for bacterial genomes to the binning procedure.}
The plot is the analogous of Fig. 1d of the main text, but with bins of different sizes. 
The binning procedure does not influence the shape of the distribution of shared components. 
}
\label{fig:Ubins}
\end{figure}

\newpage

\section{Properties of the occurrence distribution generated by an exponential frequency rank plot}

\begin{figure}[h!]
 \includegraphics[scale=1.5]{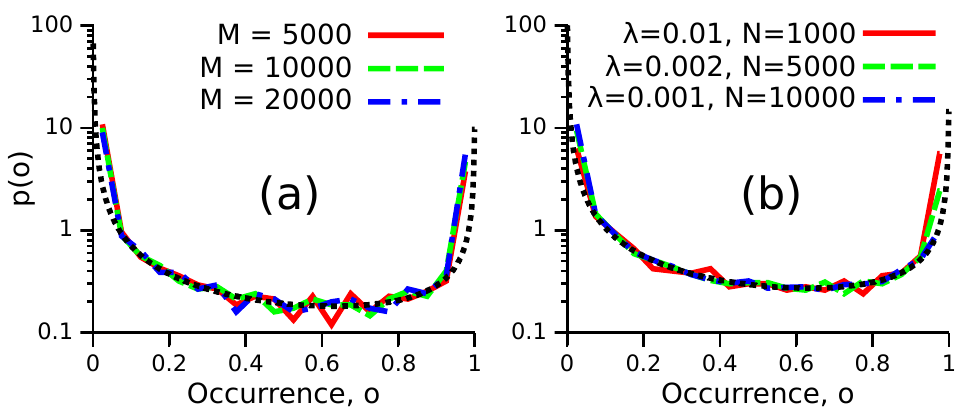}
 \caption{\textbf{Rescaling property of the occurrence distribution generated by an exponential frequency rank distribution.}
 a) The global shape of the distribution does not depend on $M$ in the limit $M \gg 1$, as shown by the Eq.~\ref{eq:largeM}. 
 The three curves are computed at fixed values of $\lambda=0.01$ and $N=1500$, and the black dotted line is the prediction of Eq.~\ref{eq:largeM}.
 The only effective parameter determining the U-shape (for $M \gg 1$) is the product $\lambda N$.
 Indeed,  panel b shows that the distribution does not change its shape while varying $N$ and $\lambda$ if their product is kept to a constant. }
 \label{fig3}
\end{figure}

The mathematical calculation described in the section IIIC of the main text can be applied to an exponential rank distribution of the form
\begin{equation} \label{eq:exp}
f_i = \frac{1}{\beta} e^{-\lambda i}, \hspace{1cm} \beta = \sum_{i=1}^{\tilde{N}} e^{-\lambda i}. 
\end{equation}
Considering a random sampling of $R$ realizations with fixed size $M$, one finds:
\begin{equation} \label{eq:exp_u}
p(o) = \frac{(1-o)^{\frac{1}{M}-1}}{\lambda M N \left( 1 - (1-o)^{\frac{1}{M}}\right)}. 
\end{equation}
Imposing the condition $M \gg 1$ this equation takes the form
\begin{equation} \label{eq:largeM}
p(o) \simeq \frac{(1-o)^{-1}}{N \lambda \log{\left[ (1-o)^{-1} \right]}}, 
\end{equation}
which provides  a good approximation for the overall distribution shape as a function of one single effective parameter $k = N \lambda$.

In the  $M \gg 1$ limit, the occurrence extreme values are $o_1 \simeq 1$ and $o_N \simeq 0$. This implies that the distribution is well defined over all  possible values of occurrence. 
Figure \ref{fig3} shows the rescaling properties of  Eq.~\ref{eq:largeM} 
by testing its independence on $M$ (panel a) and by varying $N$ and $\lambda$ while keeping  their product constant (panel b).

For rare families, one can further approximate the expression for $p(o)$ finding the expected power-law decay with exponent $-1$:
\begin{equation} \label{eq:exp_u_lim}
p(o) \simeq \frac{1}{N \lambda} o^{-1}. 
\end{equation}

We now analyze the properties of the  fraction of core components, i.e.,  those with occurrence greater than the threshold $\theta_c$. 
In order to derive the core  size one has to integrate the distribution described by Eq.~\ref{eq:exp_u} 
from $o = \theta_c$ to the maximum occurrence value $o_1$ (whose formula can be obtained from Eq. 2 of the main text).

The result reads:

\begin{equation} \label{eq:core}
 \begin{cases}
  c = 1   & \text{if} \hspace{0.3cm} o_{N} \geq \theta_c      \\
  c = - \frac{1}{N \lambda} \left( \lambda + \log{\beta} + \log{\left[ 1 - (1-\theta_c)^{\frac{1}{M}} \right]}\right) & \text{otherwise}.
 \end{cases}
\end{equation}

In the limit of  large $M$ this expression becomes 

\begin{equation}  \label{eq:core_largeM}
 c \simeq - \frac{1}{N \lambda} \left( \lambda + \log{\beta} + \log{\left[ \log{(1-\theta_c)^{-1}} \right]} -\log{M} \right), 
\end{equation}

which further simplifies  only when the logarithm of $M$ becomes dominant over the other terms.

It is worth mentioning that the expression above does not show rescaling properties, even in the regime $M \gg 1$,  and this may seem to be in contradiction with Eq.~\ref{eq:largeM}.
Nevertheless, this apparent inconsistency is basically due to the singular behavior of the occurrence distribution in $o \simeq 1$. 
In the large $M$ limit, the right boundary can be expressed as $o_1 = 1 - \epsilon$, 
where $\epsilon$ is an infinitesimal term depending on $M$ and $\lambda$, 
whose effect on the overall distribution shape is negligible (Eq.~\ref{eq:largeM}).
However, the core size is defined as the integral of the distribution. 
Therefore,  the variation of $p(o_1)$ due to a change in $M$ or $\lambda$ provides a sufficiently 
large contribution (because of the function singular behavior) which compensates the infinitesimal variation of $o_1$. 
Finally, this leads to a finite contribution to the integral and thus to the core size as it is defined in the main text. 
In general, this finite contribution has a non-trivial dependency on the parameters, 
explaining why Eq.~\ref{eq:core_largeM} does not show the rescaling property.
 
%\newpage

\section{ Difference between observed and possible vocabulary of components (Heaps' law) and its effects on the core-size estimates.}

\begin{figure}[h!]
\includegraphics[width=0.85\textwidth]{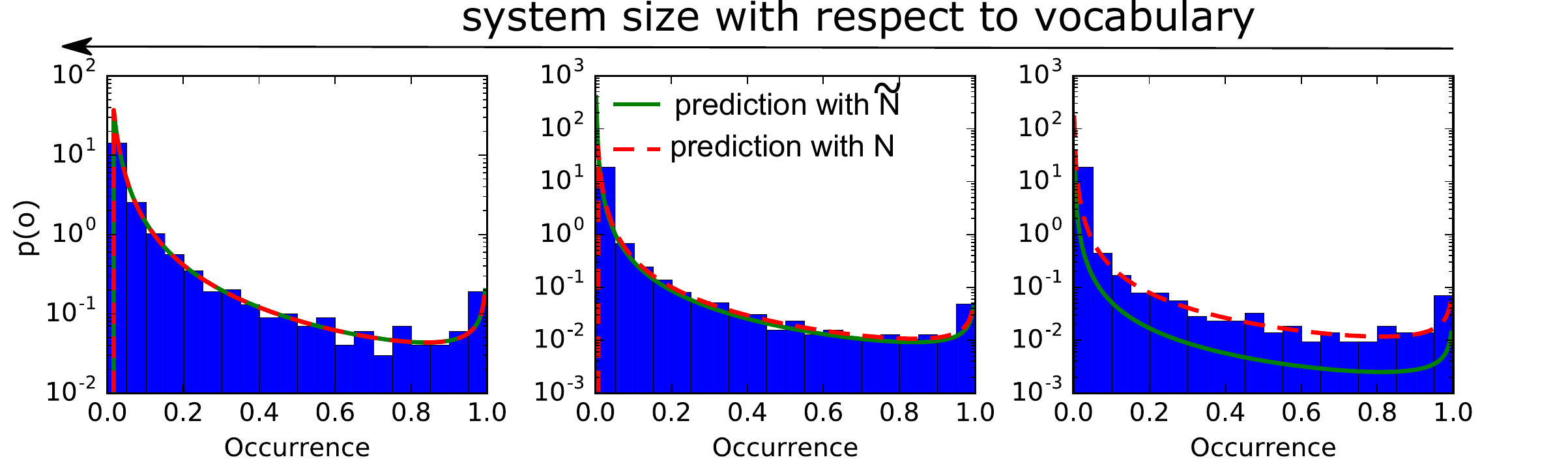}
\caption{
\textbf{Analytical prediction for the occurrence distribution of component systems with different sizes. } 
For large system sizes  $MR$ (left panel), essentially all the possible different components $\tilde{N}$ have been sampled. 
Therefore, in the analytical prediction (Eqs. 6 or 7  of the main text) for the occurrence distribution either the observed $N$ or the possible $\tilde{N}$ vocabulary can be indiscriminately used as parameters. 
On the other hand, for small systems (rigth panel) the theoretical expectation is in good agreement with numerical simulations only if the size of the actually sampled dictionary $N$ is used as a parameter.  
The parameter $\eta= \frac{MR}{N^\gamma \alpha}$, whose value gives an estimate  of the system distance from the saturation regime, is $\eta\simeq 5,1,0.1$ in the three panels respectively. 
}
\label{figSfitU}
\end{figure}

As discussed in the main text, the difference between the possible different components $\tilde{N}$ and the ones that are actually sampled in an ensemble of $R$  realizations  of size $M$ is definded by:

\begin{equation} \label{eq:heaps}
 N = \tilde{N} - \sum_{i=1}^{\tilde{N}} \left( 1 - \frac{i^{-\gamma}}{\alpha} \right)^{MR}. 
\end{equation}

This equation shows the general dependence of the  sampled dictionary on  the system size  $N(MR)$,  which is essentially a generalization of Heaps' law. 
Our analytical predictions (Eqs. 6- 9 of the main text) for the distribution of shared components have an explicit dependence on $N$ rather than on $\tilde{N}$.
As Figure~\ref{figSfitU} shows, this distinction becomes negligible in the limit of large systems, but it is in general relevant. 
A residual dependence on $\tilde{N}$ is in principle present in the normalization factor $\alpha$. However,  if $\tilde{N}$ is large (as it is the case empirically), 
this dependence is negligible and the normalization factor can actually be considered constant as it can be easily confirmed numerically.

A rough estimate of the system size at which the sampling procedure is expected to have extracted essentially all different components, thus making $N\simeq \tilde{N}$, 
can be given by introducing a crude approximation of Eq.~\ref{eq:heaps}. For large system sizes and vocabulary sizes,
the dominant term in the sum is the last term, and when this dominant term becomes negligible the sampled and the observed dictionary should roughly coincide.
The dominant term can be further approximated as 
$(1 - \frac{\tilde{N}^{-\gamma}}{\alpha})^{MR} = e^{MR ~\text{log}(1- \frac{\tilde{N}^{-\gamma}}{\alpha})} \simeq e^{-\frac{MR}{\tilde{N}^\gamma \alpha}}$. 
This approximation naturally introduces the relevant scale $\eta= \frac{MR}{\tilde{N}^\gamma \alpha}$ whose value can be used to determine if the system is close to ``saturation'',
i.e., $N\simeq \tilde{N}$ for $\eta\gg1$, or if instead a scaling analogous to Heaps' law should be expected.

% 
% \begin{figure}[h!]
% \includegraphics[width=0.85\textwidth]{figSeta.pdf}
% \caption{
% \textbf{Sampled vocabulary of components and Heaps' law. } 
% a) The sampled vocabulary $N$ approaches the total possible number of components $\tilde{N}$ as the system size $MR$ grows. 
% This is a generalization of the Heaps' law often observed in empirical component systems.
% How $N$ approaches its asymptotic value depends on a specific combination of system parameters such as the exponent $\gamma$ of the component frequency distribution 
% and the size of the possible vocabulary $\tilde{N}$. 
% b) The relevant scale $\eta= \frac{MR}{\tilde{N}^\gamma \alpha}$ is a rather  good approximation of this specific combination. In fact, the functional forms of  
%   }
% \label{figSfitU}
% \end{figure}

\begin{figure}[h!]
\includegraphics[width=0.85\textwidth]{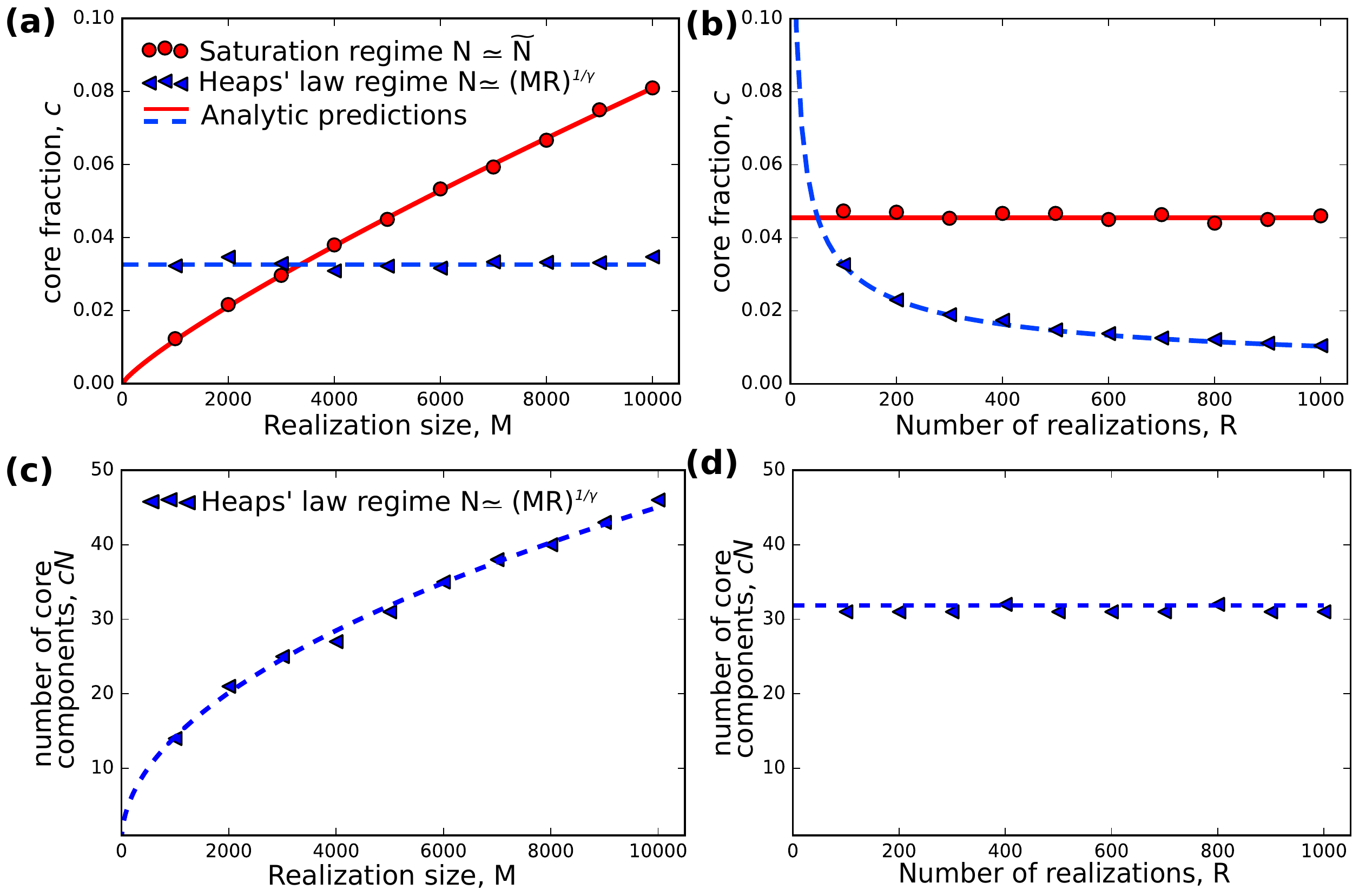}
\caption{\textbf{Core-size dependences on the number of realizations and on their typical size in different regimes.}
The fraction of core components is plotted as a function of the realization size $M$ (panel a) or of the number of realizations (panel b). 
Analytical predictions from Eqs.~\ref{eq:common_sat} and~\ref{eq:common_plaw}  are compared with numerical simulations 
of a random sampling procedure from a Zipf's law with exponent $\gamma$ in two different regimes. More specifically, as an example of the ``saturation regime'' (red circles) we used a component system with 
 $\tilde{N} = 3000$,  $\gamma = 1.2$, $R= 1000$ (in panel a), thus with  $\eta\simeq15\gg1$ for the minimal $M=10^3$,  and $M=5000$  (in panel b). 
 For the ``Heaps' law'' regime we report the illustrative example (blue triangle) of a system with
$\tilde{N} = 20000$,  $\gamma = 2$, $R=100$ (in panel a), thus with $\eta\simeq10^{-3}\ll1$ for the maximal $M=10^4$, and $M=10000$ (panel b). 
The absolute number of core components $cN$, instead of the fraction $c$, shows the  functional dependences on $M$ (panel c) and $R$ (panel d) 
described by  Eq.~\ref{eq:cN} also in the regime of parameters where Heaps' law is expected to hold. }
\label{figScore}
\end{figure}

The potential difference between the possible and the observed vocabulary of components is relevant in evaluating the depedence of the core size on the system parameters. 
For component systems with a power-law distribution of component frequencies, Eq. 12 of the main text describes the core size in terms of the rescaling parameter $k$. 
In order to translate  this general expression into the core dependences on the number $R$ and size $M$ of the realizations, different regimes have to be considered. 
If the system is close to saturation ($\eta \gg 1$)

\begin{equation} \label{eq:common_sat}
  c^{(\text{saturation})} = \left( \frac{M}{\alpha} \right)^{\frac{1}{\gamma}} \frac{1}{N} \left[ -\log(1 - \theta_c) \right]^{-\frac{1}{\gamma}} \propto \frac{M^{\frac{1}{\gamma}}}{\tilde{N}}, 
\end{equation}

where $\tilde{N}$ is a constant and this relation implies a power-law dependence 
of the core fraction on the typical realization size,  which becomes simply linear for the empirically relevant case of $\gamma=1$ (Zipf's law), 
and no dependence on the number of realizations (Figure~\ref{figScore}).

On the other hand, when $\eta$ is small, the sampled vocabulary grows sublinearly with the system size in analogy to Heaps' law. 
In the framework of a random sampling of components, this vocabulary growth can be described analytically for $\gamma > 1$ (see ref. 29 of the main text) as:

\begin{equation}
N = \Gamma \left( 1-\frac{1}{\gamma} \right) \left( \frac{M R}{\alpha} \right)^{\frac{1}{\gamma}} + o \left( \frac{M R }{\tilde{N}^{\gamma - 1}} \right).
\end{equation}

Using this expression for $N$ in the core-size estimate (Eq. 12 of the main text), we have an analytical expression of its dependencies on $M$ and $R$ 
in this ``Heaps' law'' regime: 

\begin{equation} \label{eq:common_plaw}
 c^{(\text{Heaps' law})} = \frac{R^{-\frac{1}{\gamma}}}{\Gamma\left( 1-\frac{1}{\gamma} \right)} \left[ -\log(1 - \theta_c) \right]^{-\frac{1}{\gamma}} \propto R^{-\frac{1}{\gamma}}. 
\end{equation}

In this regime the core fraction does not depend on the realization size and can be progressively reduced by adding new realizations to the ensemble as we tested numerically in Figure~\ref{figScore}.

Note that, as also discussed in the main text, 
if we consider the absolute number of core components rather than the fraction, the situation simplifies and there is no need to distinguish different regimes. 
Indeed, Eq. 12 of the main text implies that the number of core components  is simply given by 

\begin{equation}\label{eq:cN}
c N=  \left( \frac{M}{\alpha} \right)^{\frac{1}{\gamma}}  \left[ -\log(1 - \theta_c) \right]^{-\frac{1}{\gamma}}.
\end{equation}

Since $\alpha$ has only a negligible dependence on $\tilde{N}$,  the number of core components is essentially independent from $R$ for every value of $\eta$, 
and  has a power-law dependence on the typical realization size $M$ but in this case also in the ``Heaps' law'' regime.  
  This result is tested in Figure~\ref{figScore}cd. A good agreement between the analytical prediction above and numerical simulations is shown 
  in the $\eta \ll 1$ regime which indeed entails the analogous of Heaps' law.  \\

\newpage

\section{Characterization of the shape of the occurrence distribution}

\begin{figure}[h!]
\includegraphics[width=0.8\textwidth]{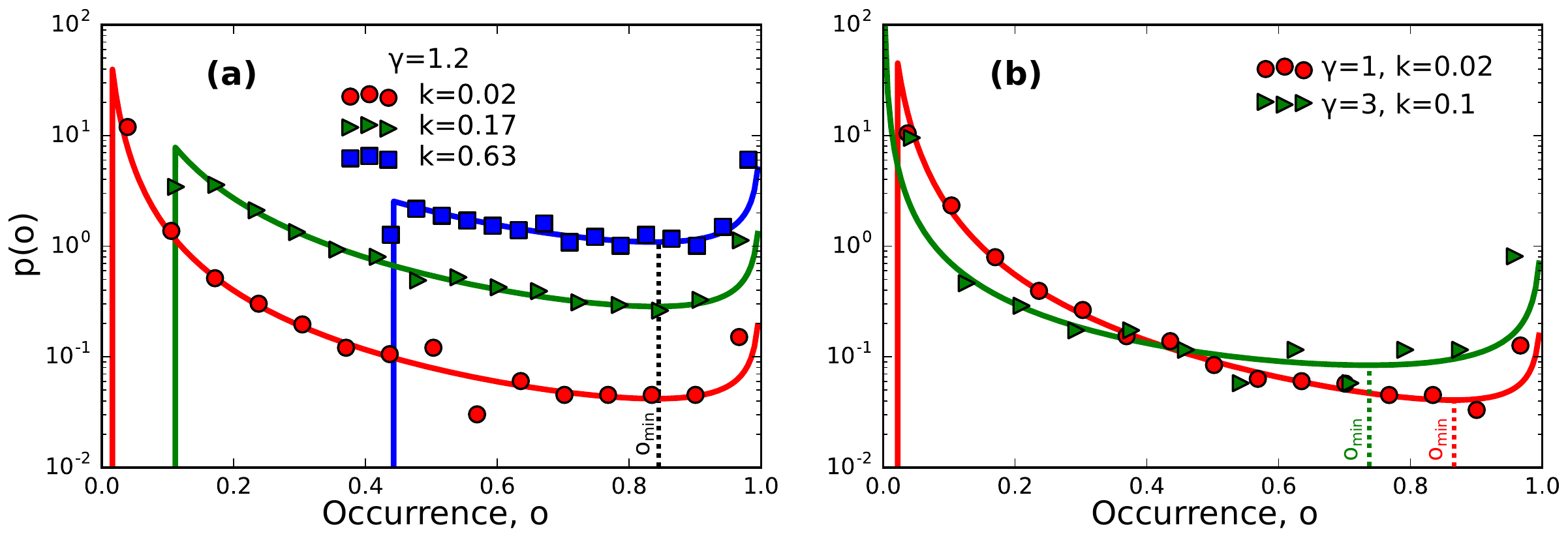}
\caption{\textbf{Shape of the occurrence distribution for different values of $\gamma$ and $k$}.
a) For a given exponent $\gamma$ of the Zipf's law, the value of the rescaling parameter $k$ determines the domain of possible occurrences and the actual fraction of core and rare components. 
b) The minimum position $o_{min}$  and the steepness of the decay of rare components are instead only a function of $\gamma$.
}
\label{fig:shape}
\end{figure}

\end{document}